\documentclass[fleqn,final]{elsart}
\usepackage{amsmath,graphicx}
\usepackage{rotating}
\begin{document}

\begin{frontmatter}

\title{Real payoffs and virtual trading in agent based market models }

\author{Fernando F. Ferreira\corauthref{Fernando}}
\ead{fagundes@ictp.trieste.it} and
\author{Matteo Marsili}
\address{International Centre for Theoretical Physics, Strada Costiera 11, Trieste, Italy 34100}
 \corauth[Fernando]{Corresponding author}
\date{\today}

\begin{abstract}
 The \$-Game was recently introduced as an extension of the
Minority Game. In this paper we compare this model with the well
know Minority Game and the Majority Game models. Due to the
inter-temporal nature of the market payoff, we introduce a two
step transaction with single and mixed group of interacting
traders. When the population is composed of two different group of
\$-traders, they show an anti-imitative behavior. However, when
they interact with  minority or majority players the \$-population
imitates the usual behavior of these players. Finally we discuss
how these models contribute to clarify the market mechanism.
\end{abstract}

\end{frontmatter}

\section{Introduction}
\label{sec:intr}
 The Minority Game (MG) \cite{challet:97-01}has addressed many issues
to understand complex systems with special emphasis on modelling
market mechanism. Although the Minority Game is a very simple model,
it contains the main ingredients to reproduce a rich behavior such as
that observed in financial market \cite{web}.  These features arise
from relatively simple rules used to define the interaction of
agents. In particular the MG assumes that market interaction is such
that it pays to be in the minority.  This assumption was recently
challenged by Andersen and Sornette \cite{sornette:01-03} who derived
a different market payoff considering the profits of round-trip
transactions (buy/sell or sell/buy). A round-trip requires at least 2
transactions and hence the market payoff depends on the price at two
different times, i.e. on the action of traders at different times. The
traders in the model introduced by Andersen and Sornette, called
\$-Game model, use this market payoff to judge the performance of
their strategies. This leads to a market behavior which is very
different from that of the Minority Game. In particular traders using
the correct market payoff are found to perform better than those who
base their behavior on the MG payoff. This models was analyzed also by
Jefferies and Johnson \cite{johnson:02-01,johnson:book}, that discuss
different aspects of market models.

A key observation is that, in the \$-game the real market payoff is
assessed on the basis of virtual trading: round trip transactions are
used to evaluate trading strategy, but they are not actually
implemented by traders. For example, if an agent bought some shares of
asset at time $t-1$, she/he will consider the virtual payoff of
selling these shares at the current price at time $t$, without
actually selling it. Using virtual trades to asses market payoff might
be a realistic behavioral assumption. Irrespective of this, a key
issue is to understand what are the consequences of this assumption on
the collective behavior of the market. More precisely, how does
market's behavior differs if agents use virtual or real trades to
asses market payoffs?

The na\"\i ve arguments which suggests that the difference is very
small, contends that prices depend very weakly on a particular
trader's transaction. Whether a transaction was really carried out or
not should not affect much the aggregate behavior. These type of
arguments are known to be problematic in heterogeneous agent market
models, and indeed fail both in the majority \cite{kozlowski} and in the
minority game \cite{CMZ} as well as in other asset market models
\cite{BMRZ}. In other words, while the impact of each trader may be
small on the aggregate, the effect of all traders neglecting their
impact can be dramatic.

We shall come to a quite similar conclusion, namely that virtual or
real trading makes a difference in market's behavior. More precisely,
we show that when agents use round-trip transactions both in the
definition of market payoff and in their actual behavior we recover a
behavior which is similar to that of the MG. Indeed round-trip
transactions induces anti-correlations in market returns and
Ref. \cite{marsili:01-01} show that market payoff reduces to those
assumed in the MG when the market returns are anti-correlated.

In order to do this we need to introduce slightly more complex
models based on two times: decision making consists on a
round-trip action where the agents decide to buy or sell at time
$t$ and do the opposite (sell or buy) at time $t+1$, then they
 evaluate the result in the next time step. In addition to this
we also aim to studying the interaction between different types of
traders, using the minority, majority or \$-Game payoff.

The paper is organized as follows: in section \ref{sec:1} we
derived the market payoff following Ref.
\cite{sornette:01-03,marsili:01-01} and we discuss the key
observable. Then we analyze single time models in section
\ref{sec:2} and then we move to two times models in section
\ref{sec:3}. Homogeneous populations and  mixed populations are
studied. Finally we give some concluding remarks in section
\ref{sec:concl}.

\section{Market Payoff}
\label{sec:1}

Now we will derive the minority and the majority mechanism from
the market's dynamics.  A single step in the Minority Game model
could be split in three stages. First of all, at time $t-\epsilon$
the agent submits an order $a_{i}(t)=\pm 1$ (buy or sell) that is
going to be useful in computing the aggregate value
$A(t)=\sum_{i}a_{i}(t)$, at time $t$. At time $t+\epsilon$ the
agents update their experience simply by evaluating the success of
their action. Focusing on one agent $i$, we can assert that if
$a_{i}(t)=1$ he/she contributes with 1\$ to the market demand and
if $a_{i}(t)=-1$ he/she contributes with $-1/p(t)$ for the supply
where $p(t)$ is the price of the cost transaction. The total
demand and supply are given by $D=\frac{N+A(t)}{2}$ and
$S=\frac{N-A(t)}{2p(t)}$ units of assets respectively. Then
following the paper \cite{marsili:01-01}, let's fix the price
$p(t+1)$ in such a way that the demand matches the supply, i.e.,
\begin{equation}
p(t+1)=p(t)\frac{N+A(t)}{N-A(t)}
\end{equation}
Let's discuss how the market payoff is defined. To reward someone
for their decisions we need to wait at least two successive
transactions. For example, if one buys a stock for the price
$p(t)$ only later, at time $t^{\prime}$ when the price will be
$p(t^{\prime})$, we are able to know if buying was profitable. So
the market payoff has an inter-temporal nature which depends at
least on two prices. We can consider for agent $i$ the capital
$C_{i}(t)=M_{i}(t)+p(t)S_{i}(t)$, which quantifies the agent
wealth at each time step. This quantity depends on the the amount
of money $M_{i}(t)$ and asset $S_{i}(t)$ which she/he owns at time
$t$, and the current price $p(t)$.

Suppose the agent $i$ decides to buy stocks $a_{i}(t)=1$ at price
$p(t+1)$, so the position changes to $M(t+1)=M(t)-1$ and
 $S(t+1)=S(t)+1/p(t+1)$. Alternatively, if the decision was to sell
 stocks, so $M(t+1)=M(t)+p(t+1)/p(t)$ and $S(t+1)=S(t)-1/p(t)$. In
 a compact way, we can express the same result as
\begin{eqnarray}
M(t+1)&=&M(t)-a(t)\frac{N-a(t)A(t)}{N-A(t)} \label{m1}\\
S(t+1)&=&S(t)+\frac{a(t)}{p(t+1)}\frac{N-a(t)A(t)}{N-A(t)}\label{s1}
\end{eqnarray}
After a simple algebra we obtain the capital increment,
\begin{equation}
C(t+1)-C(t)=[p(t+1)-p(t)]S(t)\label{cap1}
\end{equation}
which means that there is no gain from a single transaction apart
from capital gain. Then gain is due to a price changing. Now
consider that $a_{i}(t+1)=-a_{i}(t)$. This results in:
\begin{eqnarray}
M(t+2)&=&M(t)-a(t)\frac{N-a(t)A(t)}{N-A(t)} +a(t)\frac{N+a(t)A(t+1)}{N-A(t+1)}\label{m2}\\
S(t+2)&=&S(t)+\frac{a(t)}{p(t+1)}\frac{N-a(t)A(t)}{N-A(t)}-\frac{a(t)}{p(t+2)}\frac{N+a(t)A(t+1)}{N-A(t+1)}\label{s2}
\end{eqnarray}
The capital will be update by

\begin{equation}
C(t+2)=C(t)+[p(t+2)-p(t)]S(t)+Q(t+1)R(t)\frac{A(t+1)a(t)}{N}
\label{cap2}
\end{equation}
where

\begin{equation}
R(t)=\frac{1-a(t)A(t)/N}{1-A(t)/N}~~~~ and ~~~~
Q(t+1)=\frac{2}{1+A(t+1)/N}
\end{equation}
are constants of order 1 if $A(t),~A(t+1)\ll N$.

Note that in Eq. (\ref{cap2}) the first term corresponds to the
capital gain. From the second term we derive the payoff update
\begin{equation}
u_{i}^{\$}(t+1)-u_{i}^{\$}(t)=a_{i}(t)A(t+1)\label{dg}
\end{equation}
as proposed by Andersen, et.al \cite{sornette:01-03}. Previously,
Giardina, et. al \cite{giardina:01-01} proposed a quite similar
payoff function on their models. Ref. \cite{marsili:01-01}
observes that, at time $t$ the agent doesn't know the excess
demand $A(t+1)$. However he/she may resort to the expectation
given by the operator $E_{i}[\Delta u_{i}|t]$ to evaluate the best
decision to make on time $t$. Assuming that
\begin{equation}
E_{i}[A(t+1)|t]=-\phi_{i}A(t)
\end{equation}
where linearity is assumed for the sake of simplicity, the
expected payoffs increment $\Delta u_{i}(t+2)$ of agent $i$ is
\begin{equation}
E_{i}[\Delta
u_{i}(t+2)|t]=E_{i}[a_{i}(t)A(t+1)|t]=-\phi_{i}a_{i}(t)A(t)
\label{Eu}
\end{equation}
The stationary state can only occur when the expectation is
consistently validated by the dynamics as show in
\cite{marsili:01-01}, i.e., when
\begin{equation}
\phi_{i}\simeq-\frac{\langle A(t+1)A(t)\rangle}{\langle
A^{2}(t)\rangle} \label{phi}
\end{equation}
for all $i$. If $\phi_{i}>0$ the agent $i$ believes that the price
fluctuates around some equilibrium (fundamental) values $p_{f}$.
This trader is called fundamentalist and as we can see in the Eq.
(\ref{Eu}) she/he plays as a Minority Game $(<)$ player for whom
the payoff is
\begin{equation} u_{i}^{<}(t+1)-u_{i}^{<}(t)=- a_{i}(t)A(t).\label{menor}
\end{equation}

On the contrary, if the agent believes that the price is following
a trend, a movement $A(t)$ will likely be followed by fluctuations
of the same sign, i.e, $\phi_{i}<0$. These agents, called
chartists, play like in a Majority Game ($>$) and the payoff is
just
\begin{equation} u_{i}^{>}(t+1)-u_{i}^{>}(t)= a_{i}(t)A(t).\label{maior}
\end{equation}

Whether $\$$-Game is approximate by Minority Game or Majority Game
depends on the sign of $\langle A(t+1)A(t)\rangle$. We shall use
the superscript $<,>,\$ $ in the following to differentiate the
results found in the different models.

\section{Single Time Models}
\label{sec:2} In this section we will discuss three agent models
that simulate different trader's behaviors and interactions in the
stock exchange market. The set up are the same for all models,
 differing only by the payoff function that ranks their
strategies. In a generic description, consider $N$ odd agents that
buy or sell stocks, goods, currencies, etc..., to profit from the
market price fluctuation. Suppose $t$ is a discrete time. Each
trader is endowed by $\emph{S}$ quenched strategies randomly
generated at the beginning. To each strategy each trader attaches
a score $u_{i,s}(t)$ which she/he updates in the course of time.
At each time $t$, she/he undertakes the action recommended by the
highest scored strategy $s^{*}_{i}=argmax_{s} ~u_{i,s}$. The
action $a_{i,s^{*}}^{\mu(t)}(t)$ taken by agent $i$ also depend on
the value $\mu(t)$ of a public information variable.  After that,
all these orders are collected to compose the excess demand
$A(t)=\sum_{i=1}^{N} a^{\mu(t)}_{i,s^{*}_{i}}(t)$. Finally all
agents update all the payoff function $u_{i,s}$ according to Eq.
(\ref{dg}), (\ref{menor}) or (\ref{maior}) respectively for
\$-Game, Minority Game or Majority Game. The information $\mu(t)$
is updated according to the outcome $A(t)$ of the market:
\begin{equation}\mu(t+1)=\left\{
\begin{array}{c}
  2\mu(t)+1~\mod P~~~~if~A(t)>0 \\
  2\mu(t)~\mod P~~~~~~~~~if~A(t)<0 \\
\end{array}\right.
\label{mu}
\end{equation}

 Note that $A(t)=0$ never occur when N is odd. In this way
$\mu(t)$ encodes in its binary representation the signs of the
last $m$ outcomes $A(t-k)~~~k=1,...,m$ of the market. Graphically
Eq. (\ref{mu}) describes a processes on the so called de-Bruijn
graph. This is a directed graph of $2^{m}$ nodes with directed
links from nodes $k$ to nodes $2k+1~\mod~P$ and $2k~\mod~P$
\cite{Bruijn}.

\begin{figure}[h]
\begin{center}
\includegraphics[height=0.4\textheight,width=0.8\textwidth,angle=0]{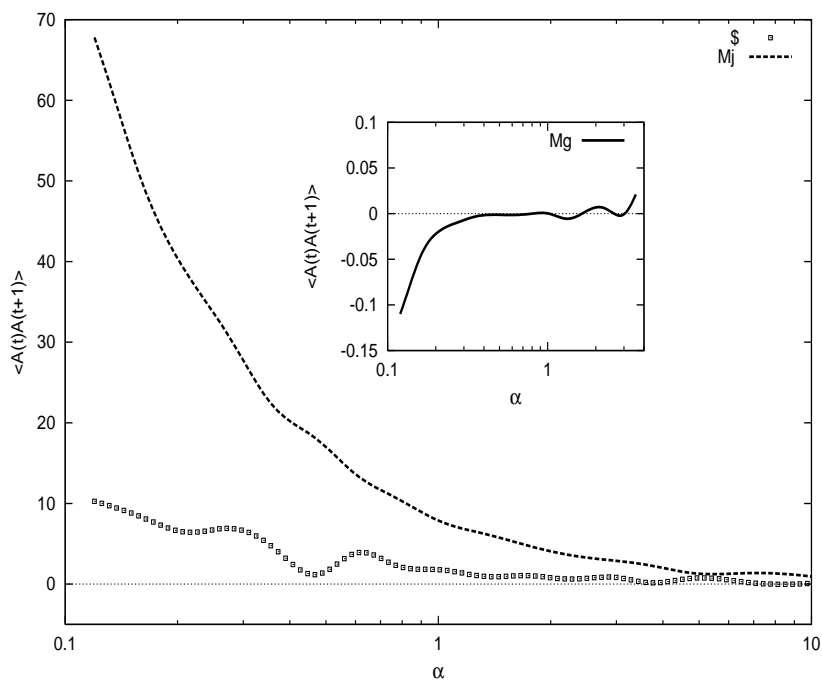}
\end{center}
\caption{Autocorrelation of the excess of demand $A(t)$ for
minority players (line inside a small box), majority players
(dashed)  and \$-players (square) in the main box. The Average
have been taken over 200 realization for all figures.} \label{fA1}
\end{figure}

\begin{figure}[h]
\begin{center}
\includegraphics[height=0.4\textheight,width=0.8\textwidth,angle=0]{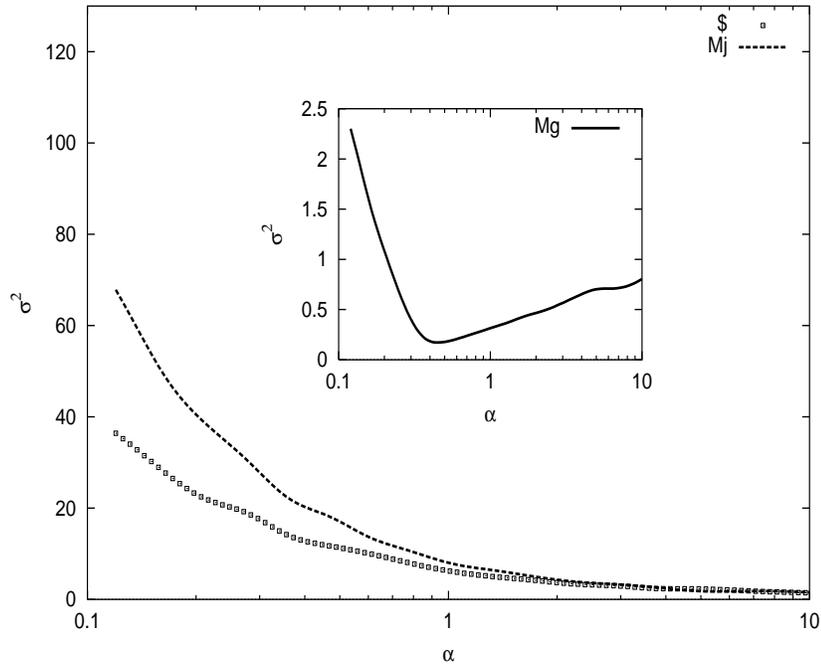}
\end{center}
\caption{Volatility of $A(t)$ for single time models. Small box
show the graph for MG. The MJ- and \$- models are plotted in the
main graph.  } \label{fsigma1}
\end{figure}

 To fix notation, we denote a temporal average of a given
time dependent quantity $W(t)$ as
\begin{equation}
\langle W \rangle = \lim_{T\rightarrow \infty}
\frac{1}{T-T_{eq}}\sum_{t=T_{eq}}^{T}W(t).
\end{equation}
 If we decompose this quantity into a conditional average on
histories, we rewrite as

\begin{equation}
\langle W|\mu \rangle = \lim_{T\rightarrow
\infty}\frac{\sum_{t=T_{eq}}^{T}W(t)\delta_{\mu(t),\mu}}{
\sum_{t=T_{eq}}^{T}\delta_{\mu(t),\mu}}
\end{equation}
Finally the average on the realization is denoted by
\begin{equation}
\bar{W}=\frac{1}{n_{r}}\sum_{i}^{n_{r}}\langle W_{i}\rangle
\end{equation}
where $n_{r}$ is the number of realization. The main free
parameter to study the stationary state of the system is
$\alpha=\frac{P}{N}$ (see \cite{savit:99-01}). Let us now analyze
the behavior of the different models. The autocorrelation is the
key to explain the others quantities for single models and to
understand the trader behavior. In the Fig. ~\ref{fA1} the
autocorrelation of the global action $A(t)$ for the Minority Game
is plotted as a function of $\alpha$ for memory $m=5$. All runs
were equilibrated for $Teq=15000$ times steps and run for 30000
times steps. We fix attention to the case of agents with $S=2$
strategies. For the Minority Game  the autocorrelation is zero for
$\alpha$ much large than the critical point $\alpha_{c}$ and the
global performance is similar to the random game. This happens
because the number of players is small compared with the amount of
information available in the market (for a discussion of Minority
Game behavior see Ref.~\cite{savit:99-01}. Increasing the number
of traders, globally they start to exploit the information and
coordinate their action until the occurrence of a phase
transition. Intrinsically minority players have a non imitative
behavior, thus the autocorrelation becomes negative below the
critical point. Conversely, the majority players are trend
follower and they induce a positive autocorrelation in $A(t)$, see
Fig. ~\ref{fA1}. What can we expect for $\$$-Game? Fig. ~\ref{fA1}
shows that the \$-Game is characterized by a positive
autocorrelation $\langle A(t)A(t+1)\rangle$. Hence we expect(see
Eq. (\ref{phi})) that the \$-Game behaves in a way which is
similar to the Majority Game. Fig.~\ref{fsigma1} supports this
conclusion by comparing, for the different models, the behavior of
$\sigma^{2}$ given by
\begin{equation}
\sigma^{2}=\frac{1}{N}\overline{\sum_{i=1}^{N}\langle A(t)^{2}
\rangle}. \label{sigma}
\end{equation}

\begin{figure}[!hb]
\begin{center}
\includegraphics[height=0.4\textheight,width=0.8\textwidth,angle=0]{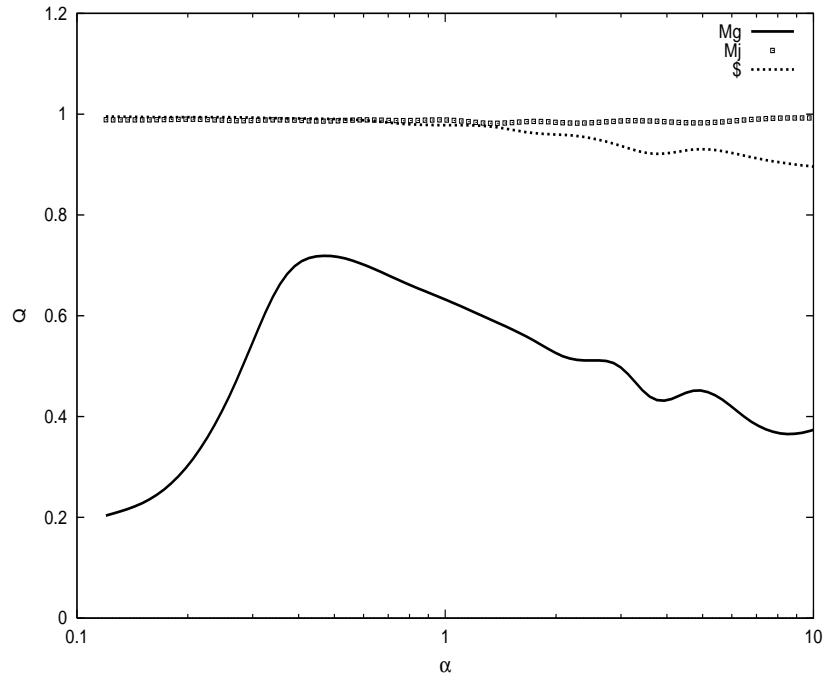}
\end{center}
\caption{Self-overlap ($Q$)of the strategies choose by plays
during the game. \$-traders and MJ-traders are frozen in the
stationary regime $Q\approx 1$. The choice of the strategies
remain fickle for MG-players $0.2<Q<0.7$ when $\alpha<0.35$}
\label{fQ1}
\end{figure}

\begin{figure}[h]
\begin{center}
\includegraphics[height=0.4\textheight,width=0.8\textwidth,angle=0]{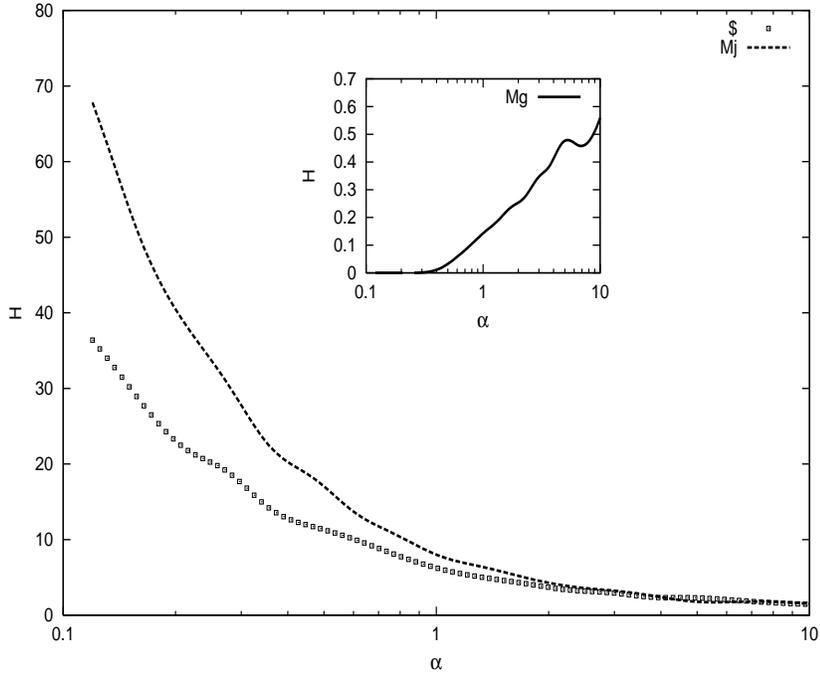}
\end{center}
\caption{Predictability $H$ (of the return $A(t)$ given the
information $\mu$). Both MJ- and \$- models are predictable. It is
due to the frozen strategies $Q\approx 1$ that leads a
quasi-deterministic dynamics in information space $\mu(t)$ (S is
very small).} \label{fH1}
\end{figure}

If you consider that $A(t)$ persist to be positive for some
period. All those strategies $s_{i}\in \emph{S}$ that is playing
with $a_{i}=1$ will be reinforce due to the payoff $\Delta
u_{i}(t+1)=a_{i}(t)A(t+1)$. Conversely, all those strategies with
$a_{i}=-1$ will be weakened. The collective tendency of the
traders is to reach a regime where they play with the same
strategy. It is very similar to what happens for MJ-Model (for
more details discussion of this models, see Ref.
\cite{kozlowski}). The \$-players persist to be in a majority
group keeping the same strategy for long. The Fig. ~\ref{fQ1},
exhibits these results clearly by reporting the self-overlap $Q$
given by:
\begin{equation}
Q=\frac{1}{N}\overline{\sum_{i=1}^{N}\langle s_{i} \rangle^{2}}.
\label{Q}
\end{equation}
This is a measure of the average fluctuation in the strategy
choice. The strategies are labelled by $s_{i}=\pm 1$, thus when
$Q=1$ it means that the players are frozen, i.e they always use
the same strategy. This is what we observe in the MJ-Model.
Fig.~\ref{fQ1} shows that this happens also in \$-Game-Model
showing that it is modelling a similar market mechanism. On the
contrary in the Minority Game we observe $Q<1$ which means that
the players are changing the strategies frequently. As a
consequence of this, the returns in the Minority Game are not
predictable. This is measured by the quantity $H$ called
predictability which is computed as
\begin{equation}
H=\frac{1}{N}\overline{\sum_{\mu}^{P}\rho^{\mu}\langle A|\mu
\rangle^{2}}.
\end{equation}
where $\rho^{\mu}$ is the probability that $\mu(t)=\mu$.
 In Fig. ~\ref{fH1} we observe that the predictability $H$ is
zero when $Q$ starts to decrease. It is known that for the
Minority Game the existence of a phase transition  for
$\alpha_{c}\approx0.34$ and the region where $H=0$ is called
symmetric phase \cite{challet:98-01}. Observe that $H=0$ implies
$\langle A|\mu\rangle=0$ for all $\mu$. In this regime is not
possible to predict the minority group based on the public
information $\mu(t)$, this property is known by economist as
\textit{no arbitrage}. For the other models (MJ and \$-Game) $H>0$
for all value of $\alpha$, that results from the fact that
$\langle A|\mu\rangle \neq 0$ for some value of $\mu$. Of course,
when the strategy of the agents are frozen the game become
predictable.

\begin{figure}[h]
\begin{center}
\includegraphics[height=0.4\textheight,width=0.8\textwidth,angle=0]{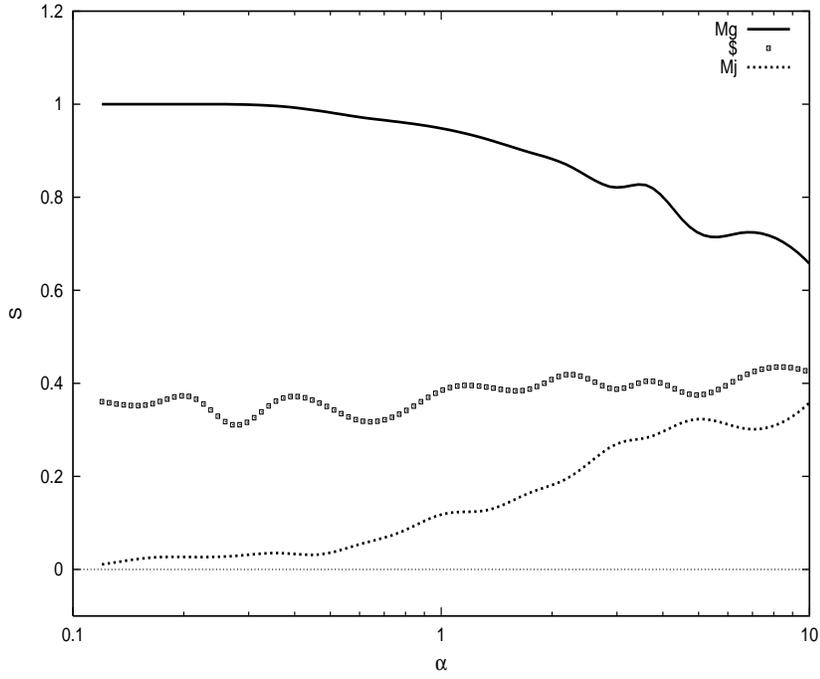}
\end{center}
\caption{Entropy $(S)$ of visited state. This quantity is
computing the dynamics of the information $\mu(t)$ for each
analyzed model, in the case of single transaction} \label{fS1}
\end{figure}

 In Fig.~\ref{fS1}, we examine the entropy for all these models given
by
\begin{equation}
S=-\frac{1}{N}\overline{\sum_{\mu}^{P}\rho^{\mu}\log_{P}(\rho^{\mu})}.
\end{equation}
 Note that $S$ was normalized to be in the interval $\in[0,1]$. This
quantity help us to analyze the dynamics of information $\mu(t)$
in the De-Bruijn graph. The MG dynamics visits all nodes in the
graph more or less uniformly. Hence $\rho^{\mu}\approx
\frac{1}{P}$ and $S$ is closed to one. In both the Majority Game
and the \$-Game we find that $S$ is smaller, this means that only
few vertices are visited. The \$-Game and MJ have a dynamics
restricted to some subgraph, i.e., only a subset of vertices
$\mu(t) \in \{0,...,P-1\}$ are visited in the stationary regime,
resulting in a small value of the entropy $S$.

\section{Two Times Models}
\label{sec:3}

Let us consider now the models whose dynamics takes place over two
time steps. Any particular agent first decides which action to
take (buy/sell) at time $t$ -- according to his/her currently best
strategy $s_i^*$ -- then takes the opposite action at time $t+1$
and finally evaluates the results according to Eq. (\ref{dg}),
(\ref{menor}) or (\ref{maior}) depending on whether he/she behaves
according to the \$, minority or majority game respectively. If
all agents behave synchronously, we just have that $A(t+1)=-A(t)$
and it is clear that the \$-Game payoffs become exactly identical
to MG payoffs. Therefore the \$-Game becomes identical with the
MG.

The non-trivial case is when agents act non-synchronously. This
means that we can classify agents into two groups, one of which
takes decisions on even times and evaluates the outcome on odd
times and the other which does the opposite. Our previous argument
suggests that when one group is much larger than the other, then
$A(t+1)\approx -A(t)$ and the game's behavior becomes that of the
MG. For this reason, we consider extreme case of equally sized
groups.

More precisely, we divide the set of agents into two sets $G_{+1}$
and $G_{-1}$ of $N/2$ agents\footnote{Now $N$ is even. Concerning
Eq. (\ref{mu}) when $A(t)=0$ we set $\mu(t+1)=2\mu(t)~\mod P$ or
$\mu(t+1)=2\mu(t)+1~\mod P$ at random, with equal probability.},
and postulate that, at time $t$, group $G_{(-1)^t}$ takes a fresh
decision whereas the group $G_{-(-1)^t}$ reacts, i.e. takes an
action which is just the opposite of the one they undertook in the
previous time step.

The interaction takes place again through the total excess demand
$A(t)$ which has now two contributions from the two groups:
\begin{equation}
A(t)= B(t)+R(t),~~~~B(t)=\sum_{i\in G_{(-1)^t}}
a_{i,s^*_i}^{\mu(t)},~~~~ R(t)=-B(t-1) \label{A2}
\end{equation}
where we introduced a different notation for the contribution
($B$) of the acting group and for that ($R$) of the reacting one.

We will first considering how agents using the different scoring
rules (\ref{dg}, \ref{menor}) and (\ref{maior}) behave when they
interact among themselves. Then we shall study the behavior of
\$-players when they are put in interaction with minority or
majority players in a mixed population.

\subsection{Single population}
\label{sec:3.1}

We analyze the situation where each agent in each of the two
different groups of traders is endowed with a set of $S=2$
strategies which are randomly drawn at the beginning.

The numerical analysis reveals that {\em i)} the behavior of a
pure population of either minority or majority players is
equivalent to that observed for a single times models {\em ii)}
the behavior of the \$-Game is similar to that of the MG.

\begin{figure}[h]
\begin{center}
\includegraphics[height=0.4\textheight,width=0.8\textwidth,angle=0]{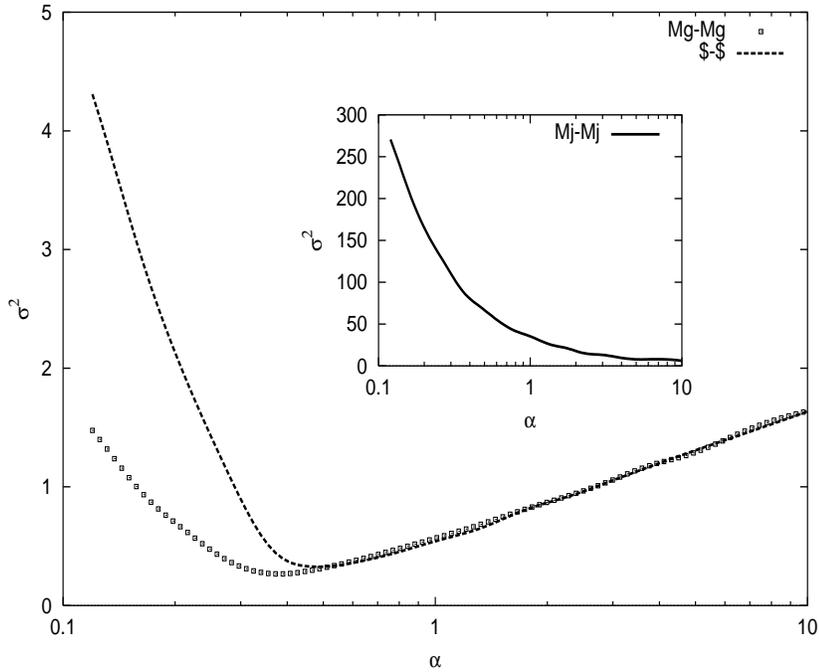}
\end{center}
\caption{Volatility in the case of two times models. The \$-\$
fluctuate more than MG-MG in the subcritical regime just because
the amplitude of the excess demand for \$-players is higher than
MG-players.} \label{fsigma2}
\end{figure}

\begin{figure}[h]
\begin{center}
\includegraphics[height=0.4\textheight,width=0.8\textwidth,angle=0]{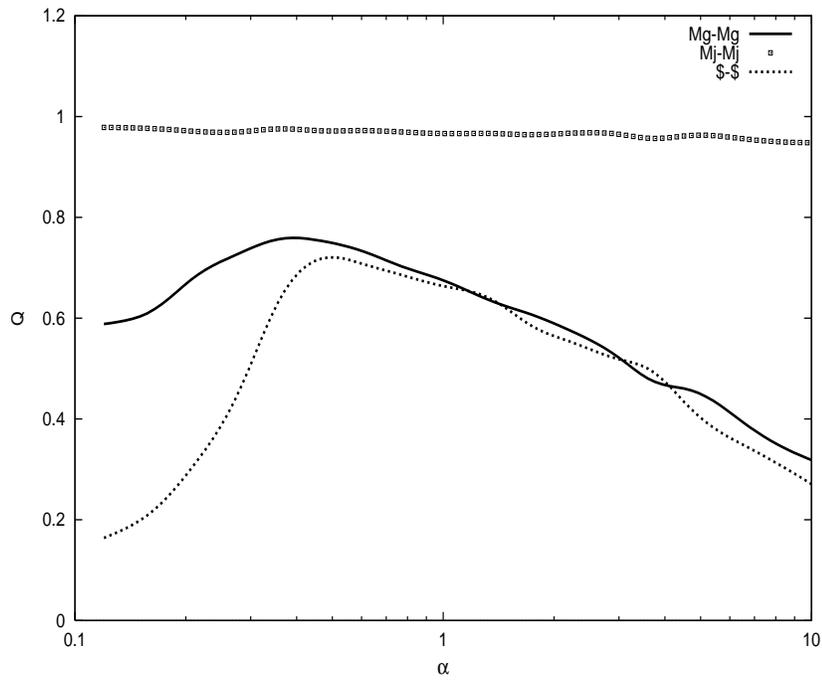}
\end{center}
\caption{The self-overlap for two times models in the homogeneous
case. Note that this quantity for MJ-MJ remain the same $Q\approx
1$. Now the \$-players choose the strategy similarly to that
observed for MG-players.} \label{fQ2}
\end{figure}

\begin{figure}[h]
\begin{center}
\includegraphics[height=0.4\textheight,width=0.8\textwidth,angle=0]{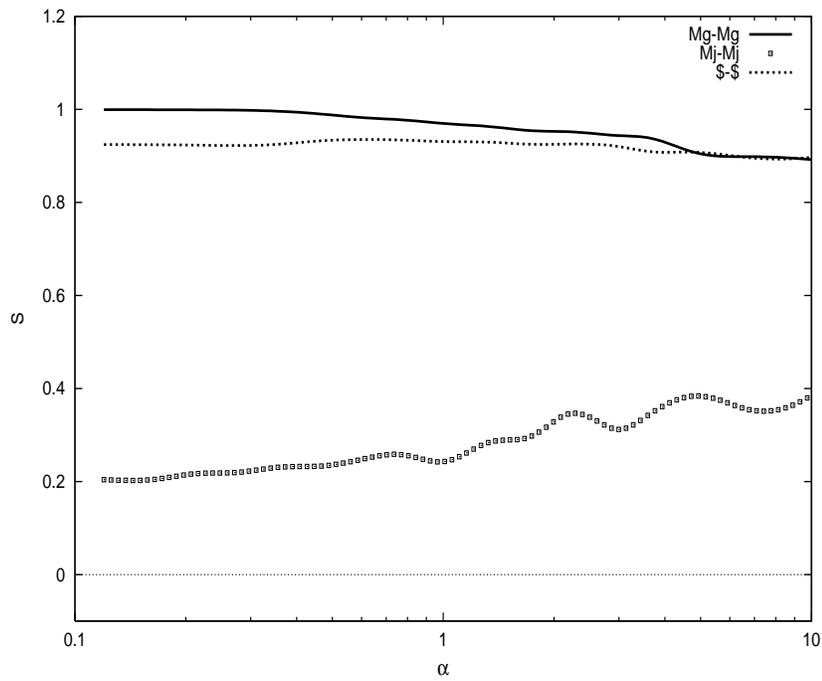}
\end{center}
\caption{Entropy for homogeneous interacting populations. The
\$-\$ model reproduce the same dynamics in the information space
as observed for MG-MG model.} \label{fS2}
\end{figure}

For example, the volatility $\sigma^{2}/N$ shows (see
Fig.~\ref{fsigma2}) the typical minimum for the MG whereas it
steadily increases as $\alpha$ decreases in the case of majority
players. In Fig ~\ref{fQ2} we see that most players get frozen
($Q\approx 1$) for MJ while the choice of the strategies remains
fickle ($0<Q^{\$,<}<0.6$) for minority players, in agreement with
the behavior of the corresponding single time models. In the case
of \$-players, both quantities ($\sigma^2$ and $Q$) show a
behavior similar to that of the MG. The same is true for the
entropy $S$, as shown in Fig. \ref{fS2}.

\begin{figure}[h]
\begin{center}
\includegraphics[height=0.4\textheight,width=0.8\textwidth,angle=0]{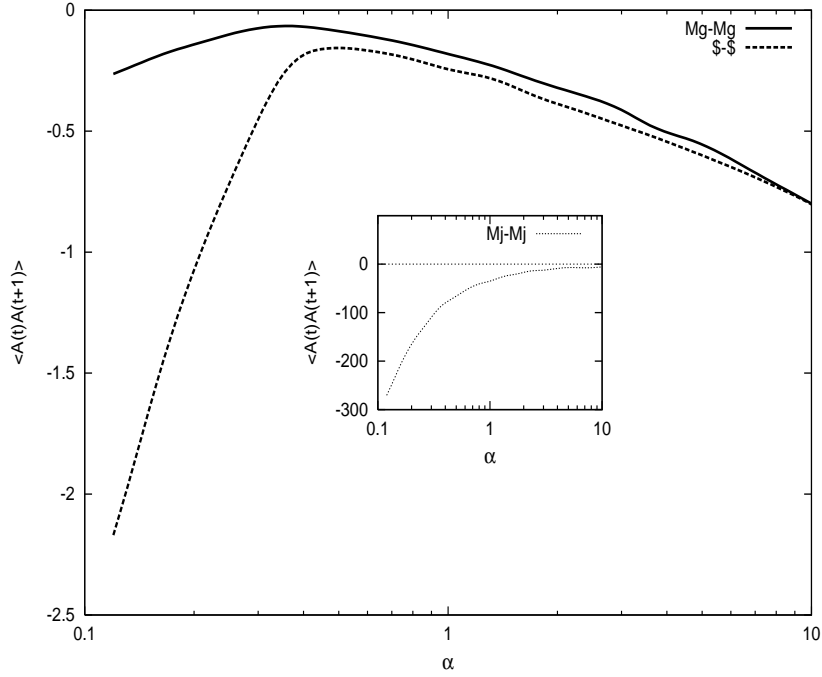}
\end{center}
\caption{Autocorrelation of $A(t)$ given by (\ref{A2}). The
negative sign is due to the interaction between populations. We
denote by MG-MG, \$-\$ and MJ-MJ the interaction between two
homogenous population composed respectively by MG, \$ and MJ. In
the small box is shown the autocorrelation for MJ-MJ model.}
\label{fA2}
\end{figure}

Indeed, for all models analyzed here, the autocorrelation of
$A(t)$ is negative, as shown in Fig ~\ref{fA2}. This is a
consequence of the alternating behavior of agents. In fact the
main contribution is due to the term $\langle
A(t)A(t+1)\rangle\simeq -\langle B^{2}(t)\rangle $ (notice that
$\langle A(t)A(t+1)\rangle$ has the same behavior as $-\sigma^2$).

\begin{figure}[h]
\begin{center}
\includegraphics[height=0.4\textheight,width=0.8\textwidth,angle=0]{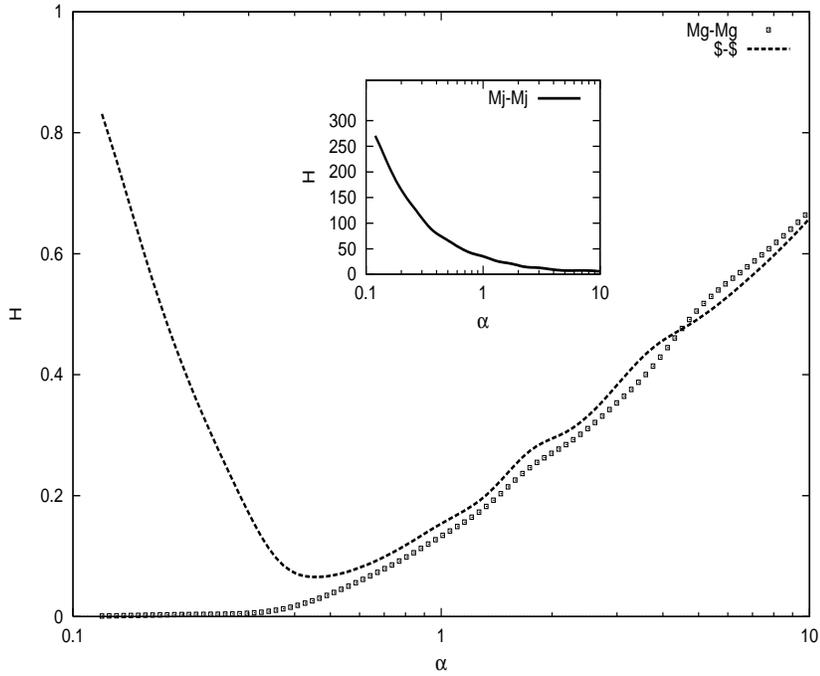}
\end{center}
\caption{The predictability $H$ for \$-\$ model quite similar for
 MG-MG model when $\alpha>0.4$, but it differs quantitatively for $\alpha<0.4$, due to the high autocorrelation
 in the reaction term $R(t)$.} \label{fH2}
\end{figure}

The only difference between \$-Game and MG appears in the behavior
of $H$. Fig.~\ref{fH2} shows that $H$ vanishes when $\alpha<0.4$
in a population of minority players (because $\langle B|\mu\rangle
\simeq -\langle R|\mu\rangle$) whereas $H$ increases for majority
players (and $\langle B|\mu\rangle \propto\langle R|\mu\rangle$).
This is in line with the behavior of single time models. The
behavior of $H$ for the \$-Game closely matches that of the MG for
$\alpha>0.4$ but it diverges from it when $\alpha<0.4$. This is a
consequence of the fact that while $\langle B|\mu\rangle \simeq
-\langle R|\mu\rangle$, (with $\langle B|\mu\rangle\neq 0$) in the
MG, we find that $\langle B|\mu\rangle$ and $\langle R|\mu\rangle$
are uncorrelated in the \$-Game. It is remarkable that when the
process $\mu(t)$ of Eq. (\ref{mu}) is replaced by random draw, as
in Ref. \cite{cavagna:99-01}, the predictability $H$ of the two
models coincides. This is because then $\langle B|\mu\rangle\simeq
0$ for all $\mu$ when $\alpha$ is small.

\subsection{Mixed population}
\label{sec:3.2}

In order to test further the behavior of \$-game players, let us
consider how they interact with minority or majority players in a
mixed population. Specifically we split each of the two groups
$G_{\pm 1}=G_{\pm 1,a}\cup G_{\pm 1,b}$ in two equal sizes (now
$N$ is a multiple of $4$). Here $a$ and $b$ label the type of
players in the sub-group. So for example $G_{\pm 1}=G_{\pm
1,\$}\cup G_{\pm 1,<}$ denotes the case where a population of
\$-players interacts with a population of minority players. Again
the interaction is on two times, but now we have two contributions
to $B(t)$ and $R(t)$ from the two sub-populations.

\begin{figure}[h]
\begin{center}
\includegraphics[height=0.4\textheight,width=0.8\textwidth,angle=0]{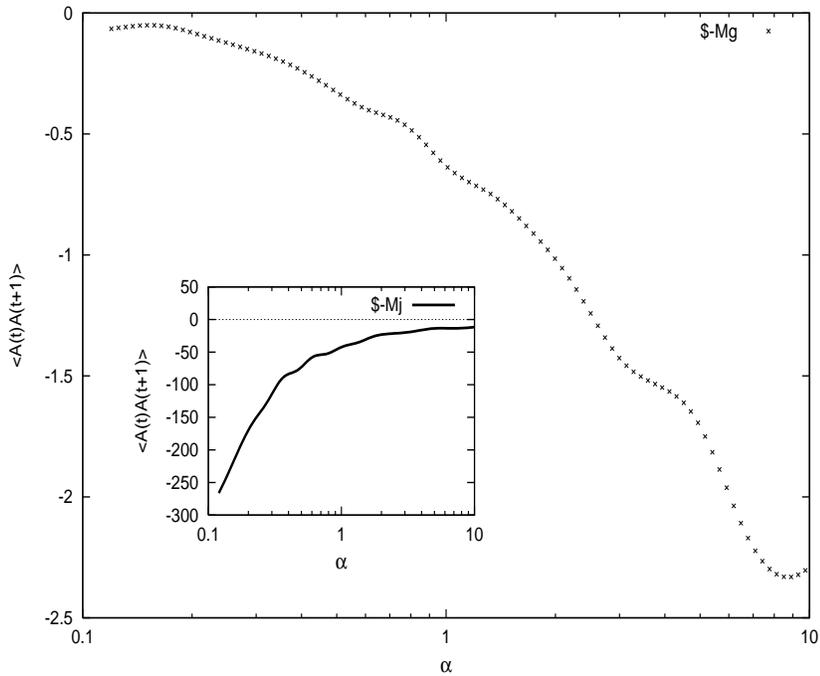}
\end{center}
\caption{The autocorrelation of $A(t)$ for \$-majority model is
shown in the small box (line).  The autocorrelation for \$-MG
interaction approximated to zero when $N$ increase.} \label{fA3}
\end{figure}

\begin{figure}[h]
\begin{center}
\includegraphics[height=0.4\textheight,width=0.8\textwidth,angle=0]{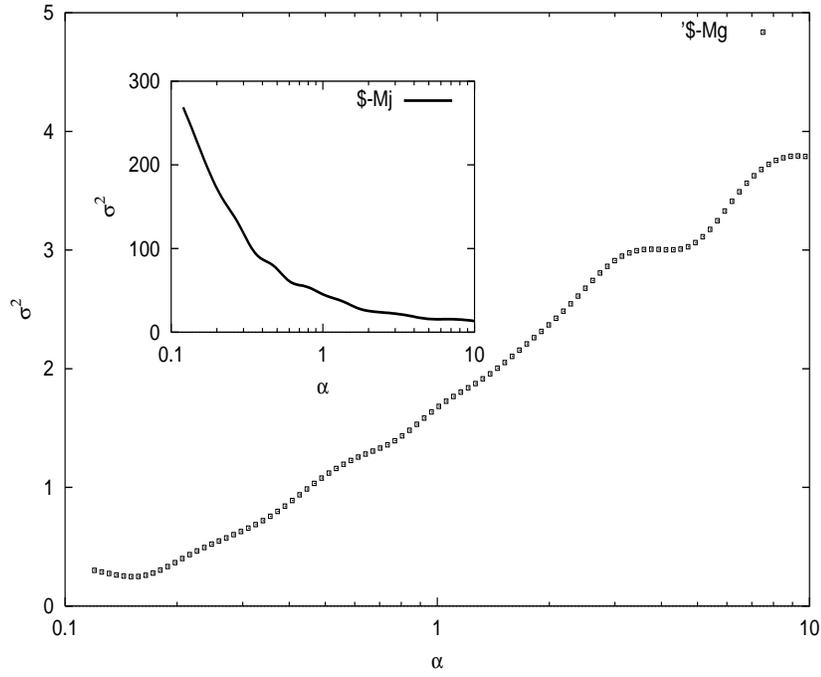}
\end{center}
\caption{The volatility $\sigma^{2}$ for mixed population in the
two times models. The \$-MG model decrease the fluctuation of the
excess demand $A(t)$ when $N$ increase.} \label{fsigma3}
\end{figure}

\begin{figure}[h]
\begin{center}
\includegraphics[height=0.4\textheight,width=0.8\textwidth,angle=0]{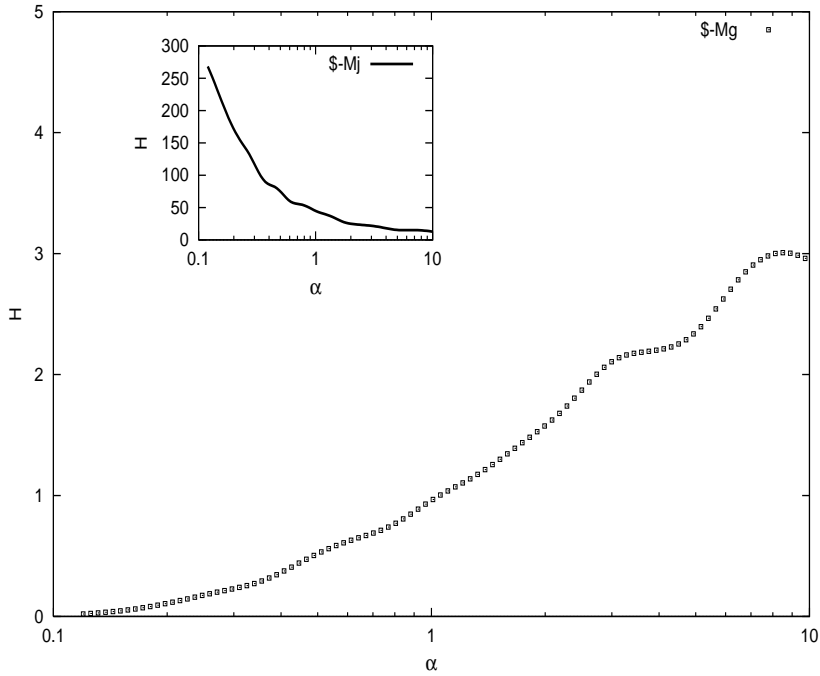}
\end{center}
\caption{Predictability $H$ for a mixed interacting population in
the two times models. Note that the sign of $A(t)$ become
unpredictable $H\approx 0$ for \$-MG model when $\alpha$
approximate to 0.1.} \label{fH3}
\end{figure}

\begin{figure}[h]
\begin{center}
\includegraphics[height=0.4\textheight,width=0.7\textwidth,angle=0]{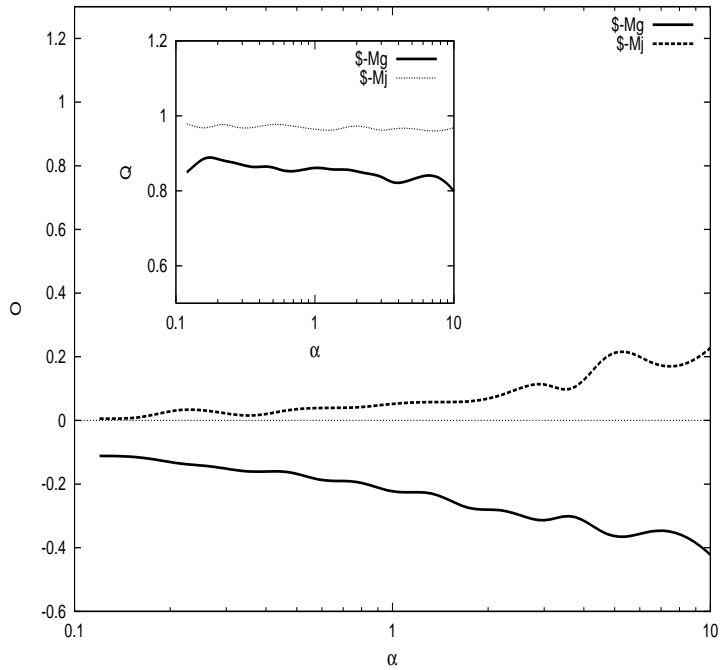}
\end{center}
\caption{The small box exhibits the self-overlap of the choice of
strategies for one fixed subpopulation. The results remains the
same for all sub-population. The quantity denoted by $O$ is
computing the overlapping of the strategy used by \$-agents
belonging to different sub-population with the same population.
The \$-traders are anti-imitative in relation to MG and imitative
when interact with MJ.} \label{fQ3}
\end{figure}

We have studied interaction of \$-players with minority or
majority players (i.e. $a=\$$ and $b=<$ or $>$). The mixed
Minority-Majority model was studied by De Martino, et.al
\cite{Martino:01-01}. We found that the autocorrelation is always
negative in both cases (see Fig. \ref{fA3}). In the case $b=<$, we
find again that the autocorrelation is proportional to
$-\sigma^2$. However contrary to the single population case, we
find that when $\alpha$ decreases the volatility $\sigma^2$
decreases (at least down to $\alpha=0.1$, see of Fig.
\ref{fsigma3}). On the contrary $\sigma^2$ increases as $\alpha$
decreases when \$-players interact with majority players (see
inset of Fig. \ref{fsigma3}). This suggests that collective
behavior is dominated by the sub-population of minority or
majority players. This is confirmed by the behavior of $H$ in Fig.
\ref{fH3}, even though it appears that the presence of \$-players
removes the phase transition (or shifts it at least to $\alpha <0.
1$). This is more clearly evident in the behavior of $Q$ in the
inset of Fig. \ref{fQ3}. When interacting with majority type
players, \$-players also get frozen ($Q\approx 1$). But also the
presence of \$-players makes minority players much less fickle
($Q\approx 0.8$). Hence \$-players increase the efficiency of
coordination of the minority players.

In order to compare the performance of agents of different types,
we give the same set of strategies to the two sub-groups $a$ and
$b$. In particular we define the overlap
\begin{equation}
O^{a,b}=\frac{1}{N} \sum \langle s^{a}_{i}\rangle \langle
 s^{b}_{i} \rangle
\end{equation}
between types $a$ and $b$. When $O=1$ ($O=-1$) the two types of
agents behave exactly in the same (opposite) way, whereas
$O\approx 0$ implies that agents of the two types behave in an
uncorrelated way. Fig. \ref{fQ3} shows that \$-players behave in a
way which is more or less uncorrelated with majority players. On
the contrary they display an anti-imitative behavior when
interacting with minority players. It seems indeed reasonable that
when minority game type of interaction prevails, as shown by the
fact that the autocorrelation of $A(t)$ is negative (see Fig.
\ref{fA3}), two agents with the same strategies would benefit from
playing in different ways.

Therefore we conclude that on one hand the collective behavior of
a mixed population ``imitates'' the behavior of the population
with which the \$-player sub-population is interacting with. But
at the microscopic level, the behavior of \$-players tend to be
loosely or negatively correlated with that of players of the other
type.

\section{Conclusions}
\label{sec:concl}
 The \$-Game model reproduces the main statistical
features observed for Majority Game whenever we consider a single
times transaction. The \$-traders freeze on the best strategy
since the payoff reinforces an imitative behavior as seen from the
positive autocorrelation of the excess of demand $A(t)$. When we
take into account the inter-temporal nature of the transaction, we
force all traders to react after all the transactions. To do so,
we proposed a model where two populations interact over two time
steps, taking decisions at different moments, out of phase. For
two interacting homogeneous populations we saw that, when traders
are all of minority or majority type, the same behavior as in
single time models obtains. A population of \$-traders shows an
anti-imitative trend in the choices of strategies behaving as
Minority Game players. Whatever the type of traders, the
autocorrelation function of the excess demand turns out to be
negative implying that the market interaction is similar to that
of the minority game.

Also in the mixed composed population, the market interaction is
of minority type, as shown by the autocorrelation of the excess
demand. The collective behavior with \$-traders interacting with a
population of minority or majority traders is similar to that of
the latter population.

In conclusion, we find that the \$-Game model exhibits a set of
new features and it poses several new and challenging questions.


\begin{thebibliography}{10}
\expandafter\ifx\csname url\endcsname\relax
  \def\url#1{\texttt{#1}}\fi
\expandafter\ifx\csname
urlprefix\endcsname\relax\def\urlprefix{URL }\fi


\bibitem{challet:97-01}
D.~Challet, Y.-C. Zhang, Emergence of cooperation and organization
in an evolutionary Game, Physica A 246 (1997) 407--418.


\bibitem{web}
Web site with a nice collection of papers and preprints
 on Econophysics: http://www.unifr.ch/econophysics.


\bibitem{sornette:01-03}
J. V.~ Andersen and D.~ Sornette, The \$-Game,Eur. Phys. J. B 31,
141-145 (2003).

\bibitem{johnson:02-01}
P. Jefferies and N. F. Johnson, Designing agent-based market
model,cond-mat/0207523.
\bibitem{johnson:book}
N. F. Johnson, P. Jefferies and P. M. Hui , Financial Market
Complexity, Oxford University Press, 2003.

\bibitem{marsili:01-01}
M. Marsili, Market mechanism and expectation in minority and
majority Games, PhysA 299 (2001) 93--103.

\bibitem{Bruijn}
http://www.scs.carleton.ca/~dquesnel/papers/debruijn/paper.html
This is a nice introdution of De Bruijn graph.

\bibitem{savit:99-01}
R. Savit et al., Adaptive Competition, Market Efficiency, Phase
Transitions, PRL, 82(10), 2203 (1999).

\bibitem{Martino:01-01}
A. De Martino, I. Giardina ~and~ G. Mosetti , Statistical
mechanics of the mixed majority-minority Game with random external
cond-mat/0305625.


\bibitem{challet:98-01}
D.~Challet, Y.~C. Zhang, On the minority Game: analytical and
numerical
  studies, Physica A 256 (1998) 514--532.

\bibitem{kozlowski}
P. Kozlowski and M. Marsili, Statistical Mechanics of Majority
Game, J. Phys. A 36 11725 (2003).

\bibitem{CMZ} D. Challet, M. Marsili, and R. Zecchina,
  Phys. Rev. Lett. {\bf 84}, 1824 (2000); M. Marsili, D. Challet, and
  R. Zecchina, Physica A {\bf 280}, 522 (2000).

\bibitem{BMRZ} J. Berg {\em et al}, Statistical mechanics of asset
markets with private information Quant. Fin. {\bf 1}
203-211  (2001).

\bibitem{cavagna:99-01}
A. Cavagna, Irrelevance of memory in the Minority Game, Phys. Rev.
E 59, R3783 (1999).

\bibitem{zhang:98-01}
Y.-C. Zhang, Modeling market mechanism with evolutionary Games,
Europhys. News
  29 (1998) 51.


\bibitem{challet:01-01}
D.~Challet, M.~Marsili, Y.-C. Zhang, Stylized facts of financial
markets and
  market crashes in minority Games, Physica A 294 (2001) 514--524.

\bibitem{lux:01-01}
T.~Lux, M.~Marchesi, Scaling and criticality in a stochastic
multi-agent model
  of financial market, Nature 397 (1999) 498--500.

\bibitem{giardina:01-01}
I.~Giardina, J.-P. Bouchaud, M.~Mezard, Microscopic models for
long ranged
  volatility correlations, Physica A 299 (2001) 28--39.

\bibitem{stanley:book:00}
H.~S. Stanley, R.~N. Mantegna, An introduction to econophysics:
correlations
  and complexity in finance, Cambridge University Press, Cambridge, 2000.

\bibitem{jefferies:00-01}
P.~Jefferies, M.~L. Hart, P.~M. Hui, N.~F. Johnson, From market
Games to real-world markets, Eur. Phys. J. B 20 (2001) 493--501.




\end{thebibliography}

\begin{figure}[h]
\begin{center}
\includegraphics[height=0.35\textheight,width=0.8\textwidth,angle=0]{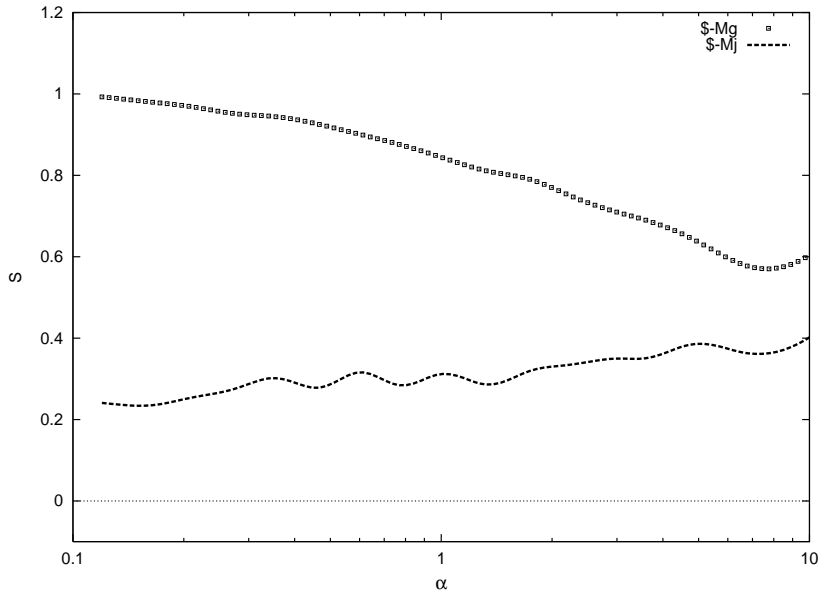}
\end{center}
\caption{The entropy $S$ show that the dynamics of information for
mixed population (\$-MG and \$-MJ) are qualitatively similar to
that observed for MG and MJ (single times models)} \label{fS3}
\end{figure}

\end{document}